\documentstyle[12pt,aasms4,epsfig]{article}
\def\paren#1{\left( #1 \right)}
\def\brace#1{\left\{ #1 \right\}}
\def\bra#1{\left[ #1 \right]}
\def\angle#1{\left\langle #1 \right\rangle}

\begin{document}
\lefthead{Shiho, Piran \& Sari}
\righthead{}
\title{Hydrodynamics of a Relativistic Fireball: the Complete Evolution}
\author{Shiho Kobayashi\authoremail{shiho@alf.fiz.huji.ac.il},
        Tsvi Piran\authoremail{tsvi@shemesh.fiz.huji.ac.il} and  
        Re'em Sari\authoremail{sari@shemesh.fiz.huji.ac.il}} 
\affil{Racah Institute of Physics, The Hebrew University, 
       Jerusalem, 91904 Israel}

%%%%%%%%%%%%%%%%%%%%%%%%%%%%%%%%%%%%%%%%%%%%%%%%%%%%%%%%%%%
\begin{abstract}
  We study numerically the evolution of an adiabatic relativistic
  fireball expanding into a cold uniform medium. We follow the stages
  of initial free expansion and acceleration, coasting and then
  deceleration and slowing down to a non-relativistic velocity. We
  compare the numerical results with simplified analytical
  estimates. We show that the relativistic self similar
  Blandford-McKee solution describes well the relativistic
  deceleration epoch. It is an excellent approximation throughout the
  relativistic deceleration stage, down to $\gamma \sim 5$, and a
  reasonable approximation even down to $\gamma \sim 2$ though the
  solution is rigorous only for $\gamma \gg 1$. We examine the
  transition into the Blandford-McKee solution, and the transition
  from the solution to the non-relativistic self similar Sedov-Taylor
  solution. These simulations demonstrate the attractive nature of the
  Blandford-McKee solution and its stability to radial perturbations.
\end{abstract}
%%%%%%%%%%%%%%%%%%%%%%%%%%%%%%%%%%%%%%%%%%%%%%%%%%%%%%%%%%%%%%
\keywords{hydrodynamics ---
          relativity ---
          shock waves ---
	  gamma-ray bursts}
%%%%%%%%%%%%%%%%%%%%%%%%%%%%%%%%%%%%%%%%%%%%%%%%%%%%%%%%%%%%%%
\section{Introduction}

Sedov (1946), Taylor (1950) and Von Neumann (1947) discovered, in the
forties, a self similar solution of the strong explosion problem, in
which a large amount of energy is released on a short time scale in a
small volume. This solution is known today as the ``Sedov-Taylor''
self similar solution.  It describes a shock wave propagating into a
uniform density surrounding. The shock wave and the matter behind it
decelerate as more and more mass is collected. This solution describes
well the adiabatic stage of a supernova remnant evolution.

Blandford and McKee (1977) have later established a self similar
solution describing the extreme relativistic version of the strong
explosion problem. In this solution the Lorentz factor of the shock
and the fluid behind it is much larger than unity. Such high Lorentz
factors arise if the rest mass contained within the region where the
energy $E$ was released is much smaller than $E$. In other words when
the region containing the energy is ``radiation dominated'' rather
than matter dominated. Such a region was later termed a ``fireball''.

Cavallo and Rees (1978) have considered the physical processes
relevant in a radiation dominated fireball as a model for gamma-ray
bursts (GRBs). Goodman (1986) and Paczy\'nski (1986) have then
considered the evolution of such a fireball. They have shown that a
initially the radiation-pair plasma in a purely radiative fireball
behaves like a fluid and it expands and accelerated under its own
pressure until the local temperature drops to $\sim 20$keV, when the
last pairs annihilate and the fireball becomes optically thin.  Later
Shemi and Piran (1990) considered matter contaminated fireballs. They
have shown that, quite generally, all the initial energy will be
transfered to the baryons in such fireballs whose final outcome is a
shell of relativistic freely expanding baryons.  Piran, Shemi and
Narayan (1993) and M\'esaz\'aros Laguna and Rees (1993) have later
carried these calculations in greater details. These works show that
an initially homogeneous fireball will first accelerate while
expanding and then coast freely as all its internal energy was
transformed to kinetic energy.

The surrounding matter will eventually influence the fireball after
enough external matter has been collected and most of the energy has
been transfered from the shell to the ISM.  If the surrounding matter
is diluted enough, then this will take place only after the initial
free acceleration phase.  This influence was considered by
M\'esaz\'aros and Rees (1992), and Katz (1993), who suggested that the
GRB is produced during this stage. The detailed shock evolution was
later studies by Sari and Piran (1995). It is only after these stages,
that the fireball have given the ISM most of its energy and then the
self similar deceleration solution of Blandford-McKee applies. When
the shock decelerates enough so that it is no longer relativistic, it
is described by the Sedov-Taylor solution.

Today it is widely accepted that GRBs involve relativistic expanding
matter of this kind. While the GRB itself is produced, most likely,
via internal shocks (Narayan, Paczy\'nski \& Piran, 1992; Rees \&
M\'esaz\'aros, 1994; Sari \& Piran, 1997).  The observed GRB
afterglow corresponds, on the other hand to the slowing down of this
relativistic flow. This has led to an increasing interest in the
fireball solution and in its various regimes. In this paper we study
the evolution of a homogeneous fireball focusing on its interaction
with the surrounding matter.  We do not consider here internal shocks,
which arise due to interaction within the relativistic flow and
require nonuniform velocity.

We have developed a spherically symmetric relativistic Lagrangian code
based on a second order Gudnov method with an exact Riemann solver to
solve the ultra-relativistic hydrodynamics problem. With this code, it
is possible to track the full hydrodynamical evolution of the fireball
within a single computation, from its initial ``radiation dominated''
stage at rest through its acceleration, coasting, and shock formation,
up to the relativistic deceleration and finally to the Newtonian
deceleration.  With typical parameters this computation spans more
than eight orders of magnitudes in the size of the fireball and more
than twenty orders of magnitudes in its density. We describe these
computations here.  We show that, quite generically, the solution
converges during the relativistic deceleration phase to the
Blandford-McKee solution and then it transforms to the Sedov-Taylor
solution. Even though the attractive nature of the Blandford-McKee
solution suggests that it is stable we explore explicitly the
stability of this solution and we show that it is stable to radial
perturbations.

We review, first, in section \ref{sec_Analytic} the current analytic
understanding of the fireball evolution thorough the following stages:
(i) free acceleration and coasting, (ii) energy transfer (iii)
relativistic self similar solution and (iv) Newtonian Sedov-Taylor
solution.  We discuss the numerical results in section
\ref{numerical}.  In section \ref{perturbation} we examine the
evolution of perturbations to the Blandford-McKee solution. We discuss
the implications of these results in section \ref{conclusions}.
%%%%%%%%%%%%%%%%%%%%%%%%%%%%%%%%%%%%%%%%%%%%%%%%%%%%%%%%%%%%%%
\section{Analytic Estimates}
\label{sec_Analytic}
The evolution of a fireball is characterized by several phases.
The transitions between these phases are determined by several
critical radii that are summarized in table \ref{t:radii}.

\begin{center}
\begin{table*}[ht!]
\caption{\it Critical Radii}
\label{t:radii}

  \begin{center}
\begin {tabular}{|c|c|c|}
  \hline
  $R_0$      & Initial Fireball Size              &              \\
  $R_L$      & Coasting          &$R_0  \eta$   \\
  $R_s$      & Spreading (and Internal Shocks)    &$R_0  \eta^2$ \\
  $R_\Delta$ & External Shocks (the NRS Case)  &$ l^{3/4} \Delta^{1/4}$ \\
  $R_\gamma$ & External Shocks (the RRS Case)  &$l/\eta^{2/3}$\\
  $R_N$      & Relativistic Reverse Shock      &
   $l^{3/2}/\Delta^{1/2}\eta^2$\\
  $l$        & Sedov Length                    &$(E/ \rho_1)^{1/3}$ \\
  \hline
\end{tabular}
\end{center}
\end{table*}
\end{center}
%%%%%%%
\subsection{Free Acceleration and Coasting}
We consider a homogeneous fireball of energy $E$ and a baryonic load of
total mass $M_0$ confined initially in a sphere of radius $R_0$. We 
define the dimensionless entropy (or the initial random 
Lorentz factor) $\eta \equiv E/M_0$. This fireball expands into a  
surrounding low density medium (with a density $\rho_1$) which we will 
refer to as the ISM. This can be considered to be a free expansion in
its initial stage. After a short acceleration phase, the motion
becomes highly relativistic. Conservations of baryon number,
energy and momentum yield  the following conservation laws along
a null flow line of each fluid element in the shell (Piran, Shemi \&
Narayan, 1993):
\begin{equation}
r^2 \rho \gamma, \ \ \   
r^2 p^{3/4} \gamma \ \ \  \mbox{and} \ \ \ 
r^2 (4p+\rho) \gamma^2 = \mbox{constant},
\label{eq:linear}
\end{equation}
where $r(t)$, $\gamma(t)$, $p(t)$ and $\rho(t)$ are the radius,
Lorentz factor, pressure and rest mass density of the fluid element,
respectively. In this paper, distance, time, velocity and the
corresponding Lorentz factors are measured in the observer
frame. Thermodynamic quantities ($p$ and $\rho$) are measured in the
local fluid frame. We use units in which the speed of light $c=1$. The
above equations assume an adiabatic gas index of $4/3$.

Initially the fireball is extremely hot ($p \gg \rho$), so that 
equation (\ref{eq:linear}) yields
\begin{equation}
\gamma \propto r, \ \ \ \rho \propto r^{-3} 
\ \ \ \mbox{and} \ \ \ p \propto r^{-4},
\label{eq:scaling_exp}
\end{equation}
(Goodman, 1986; Paczy\'nski, 1986; Shemi \& Piran, 1990). The fireball
is approximately homogeneous in the local frame, but due to
relativistic effects it appears it appears to an observer at rest as a
narrow shell with a radial width $\Delta \sim r/\gamma \sim R_0$
(Shemi \& Piran, 1990; Piran, Shemi \& Narayan, 1993). As the fireball
expands, the internal energy is converted into kinetic energy of the
baryons. At the radius $R_L \equiv \eta R_0$, the fireball uses up all
the internal energy and the approximation $p \gg \rho$ breaks
down. This is the end of the acceleration phase.

Now, the internal energy of the fireball becomes negligible compared
to the rest mass energy ($p \ll \rho$), and equation
(\ref{eq:linear}) yields
\begin{equation}
\gamma = \mbox{constant}, \ \ \ \rho \propto r^{-2} \ \ \ 
\mbox{and} \ \ \ p \propto r^{-8/3},
\label{eq:scaling_coasting} 
\end{equation}
(Piran Shemi \& Narayan, 1993). The fireball behaves like a pulse of
energy with a frozen radial profile propagating at almost the speed 
of light.

%%%%%%%%
\subsection{Spreading, The Reverse Shock and Energy Transfer}
This frozen pulse approximation on which equation (\ref{eq:linear}) is
based breaks down ultimately at the radius $R_s \equiv R_0\eta^2$.
Each fluid shell moves with a slightly different velocity and the
fireball begins to spread at $R_s$.  Internal shocks will take place
around $R_s$ if the fireball is inhomogeneous and the velocity is not
a monotonic function of the radius.  As mentioned earlier, these
shocks produce, most likely, the observed GRB. However, even under
optimal condition they cannot convert more than about a quarter of the
kinetic energy to radiation. Hence even an inhomogeneous shell will
continue carrying ample kinetic energy beyond this stage. We consider
here only homogeneous fireball. The  possible effect of
internal shocks on the fireball evolution has been discussed extensively 
in another papers (Kobayashi, Piran \& Sari, 1997; 
Daigne \& Mochkovitch, 1997)

The coasting can also end if the surrounding matter begins to
influence the shell. The interaction between the shell and the ISM can
be described by two shocks: a forward shock propagating into the ISM
and a reverse shock propagating into the shell. Sari and Piran (1995)
have defined three critical  radii in this respect: $R_{N}\equiv
l^{3/2}/{\Delta}^{1/2}\eta^2$ where the energy density produced by the
shocks becomes high enough so that the reverse shock is relativistic
and begins to reduce the Lorentz factor of the shell considerably;
$R_{\Delta}\equiv l^{3/4}{\Delta}^{1/4}$ where the reverse shock
crosses the shell; and $R_{\gamma}\equiv l/\eta^{2/3}$ where the mass
of the shocked ISM is $M_0/\eta$.  Here $l\equiv (E/\rho_1)^{1/3}$ is
the Sedov length. Fortunately, a simple relations between these four
radii can be given in terms of the dimensionless variable $\xi\equiv
(l/\Delta)^{1/2}\eta^{-4/3}$;
\begin{equation}
\xi^2R_s=\sqrt{\xi}R_{\Delta}=R_{\gamma}=R_N/\xi.
\end{equation}

If initially $\xi>1$ then $R_s$ is the smallest radius. The shell
begins to spread at $R_s$. After that the width $\Delta$ satisfies
$\Delta=r/\gamma^2 \propto r$ and the scaling of the shell parameters
becomes
\begin{equation}
\gamma = \mbox{constant}, \ \ \ \rho \propto r^{-3} \ \ \ 
\mbox{and} \ \ \ p \propto r^{-4}.
\end{equation}
During the spreading phase the value of $\xi > 1$ decreases. However,
as long as $\xi>1$ the relation $R_s < R_\Delta<R_\gamma<R_N$ is
valid. When $\xi \sim 1$ these different radii become comparable: the
reverse shock crosses the shell, it becomes mildly relativistic and an
ISM mass of $M_0/\eta$ was collected. Since the reverse shock is just
mildly relativistic at this stage, the shell's Lorentz factor have
changed only by a factor of order unity. We call this the Newtonian
Reverse Shock (NRS) case since the reverse shock is Newtonian relative
to the unshocked shell. This is not to be confused with the fact that
$\gamma \gg 1$ in this case and the forward shock is
ultra-relativistic.

If initially $\xi<1$ then $R_N$ is the smallest radius. The reverse 
shock  becomes relativistic before it crosses the shell. 
At $r>R_N$ the reverse shock begins to reduce considerably the Lorentz
factor of the shell's matter that it crosses. The Lorentz factor 
of the shocked material is 
\begin{equation}
\gamma(r)=l^{3/4}\Delta^{-1/4}r^{-1/2}
\label{eq:gammacd}
\end{equation}
(Sari, 1997). The shell has decelerated significantly by the time
that the reverse shock has crossed the shell at $R_\Delta$. At this
stage, the Lorentz factor was reduced from its initial value of $\eta$
to $\eta
\xi^{3/4}=(l/\Delta)^{3/8} \ll\eta$. We call this the Relativistic
Reverse Shock (RRS) case as most of the deceleration is done by a
strong relativistic reverse shock.

\begin{figure}[b!] % fig 1
\centerline{\epsfig{file=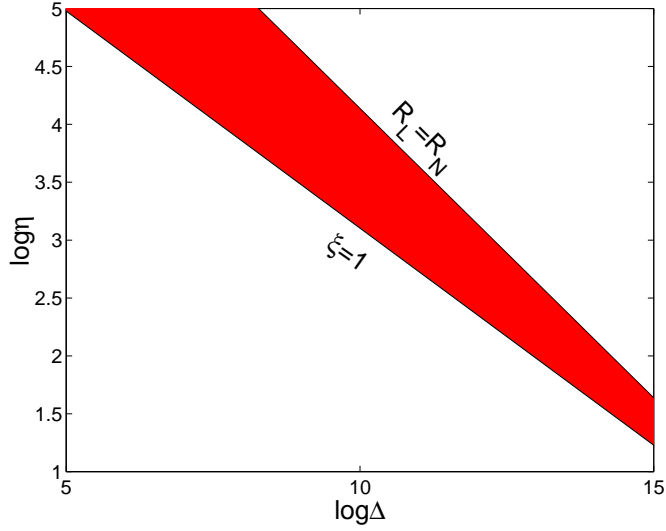,width=3.5in}}
\vspace{10pt}
\caption{Allowed parameter region for $l=1.9\times10^{18}$cm
($E=10^{52}$ergs and $\rho_1=1$proton$/cm^3$). The line ($\xi=1$)
separates the NRS case (lower left) and the RRS case (upper right).
$R_L<R_N$ is the lower left region of the line ($R_L=R_N$).
}
\label{fig:delta_eta}
\end{figure}

The coasting radius, when the shell begins to coast freely, $R_L$, is
related to the other radii by a simple relation, $\eta
\xi^2R_L=\xi^2R_s=\sqrt{\xi}R_\Delta=R_\gamma=R_N/\xi$.  In
the NRS case $\xi > 1$ and $R_L$ is the smallest radius.  All the
initial thermal energy is converted to kinetic energy before the shell
begins to decelerate.  In the RRS case ($\xi < 1$) it is possible that
the deceleration, due to external matter, begins before all the
energy of the fireball was converted to kinetic energy.  The
conditions $R_L < R_N$ and $\xi < 1$ yield $(l/\Delta)^{3/8}<\eta <
\sqrt{l/\Delta} $ (see figure \ref{fig:delta_eta}):  
$\xi$ should be larger than $(\Delta/l)^{1/6}$ in order that $R_L<
R_N$.  We limit the discussion to this case here,$R_L < R_N$, i.e.,
the fireball transforms all its internal energy to kinetic energy
before the ISM begins influence its evolution.

The NRS case is relevant if the initial fireball is small
and contains relatively many baryons (equivalently, $\eta$ is 
relatively small), while the RRS case is relevant if
the fireball is large and it is less polluted by baryon. 
The shell has given the ISM most of its energy either at
$R_\gamma (=R_N=R_\Delta)$ in the NRS case, or at $R_\Delta$ 
in the RRS case, 

%%%%%%%%
\subsection{Relativistic Self Similar Deceleration}
For $r>R_{\gamma}$ for the NRS case, or $r>R_{\Delta}$ for the RRS
case, the shocked ISM contains most of the energy $E$.  From this
stage the shell plays a negligible part in the consecutive
evolution. The profile of the shocked ISM is determined now just by
two parameters, $E$ and $\rho_1$.

When the forward shock reaches a radius $R$ the fireball has a mass $4
\pi \rho_1 R^3/3$. The energy in the shocked fluid is therefor
$\propto \rho_1 R^3\gamma^2$. Since this equals the total energy of
the system $E$, we obtain a scaling law, $\gamma \propto
(E/\rho_1)^{1/2} R^{-3/2}$. The exact proportionality constant depends
on the profiles behind the shock. Blandford and McKee (1976) describe
an analytic self similar solution, in which the Lorentz factor, the
density and the pressure are given by
\begin{equation}
\gamma(t,r)=\frac{1}{\sqrt{2}}\Gamma\chi^{-1/2}, \ \ \ \\
\rho(t,r)=2\sqrt{2}\rho_1 \Gamma\chi^{-5/4}, \ \ \ \\
p(t,r)=\frac{2}{3}\rho_1\Gamma^2\chi^{-17/12},
\label{eq:bm}
\end{equation}
where $\Gamma(t) \equiv \sqrt{17/8\pi}(R/l)^{-3/2}$ is the Lorentz
factor of the shock itself, and the similarity variable $\chi$ is
defined by $\chi(t,r) \equiv 1+8\Gamma^2(1-r/R)$. The shock radius
is given by $R=R(t) \sim (1-1/8\Gamma^2)t$. In the previous section
$r$ denotes the radius of a fluid element in the shell, but $r$ and 
$t$ are independent coordinates here.
%%%%%%%%
\subsection{The Sedov-Taylor Solution}

The Blandford-McKee self similar solution is derived with the
assumption $\gamma,\Gamma \gg 1$. This assumption breaks down when the
shock sweeps out a volume $\sim l^3$ of ISM and its motion becomes
non-relativistic. At the radius $l$ the Sedov-Taylor (Sedov, 1946; Von
Neumann, 1947; Taylor, 1950) solution becomes a good approximation .
A characteristic length scale at a time $t$, which we can form from
the two parameters $E$ and $\rho_1$, gives the shock radius
$R(t)\equiv\alpha (Et^2/\rho_1)^{1/5}$ within a numerical constant
factor $\alpha$ depending on the adiabatic constant $\hat{\gamma}$
($\alpha=0.99$ for $\hat{\gamma}=4/3$). The velocity of the shock is
$u\equiv dR/dt=2R/5t$. The velocity $\beta_2$, the density $\rho_2$
and the pressure $p_2$ just behind the shock can be expressed in terms
of $u$: $\beta_2=6u/7, \rho_2=7\rho_1, p_2=6\rho_1u^2/7$ for the
adiabatic index $\hat{\gamma}=4/3$.  The profile throughout the region
behind the shock is given by the velocity $\beta=2rV/5t$, the density
$\rho=\rho_1 G$ and the pressure $p=3r^2\rho_1GZ/25t^2$ (Landau \&
Lifshitz, 1987). The dimensionless variables $V, G, Z$ are functions
of the similarity variable $\zeta
\equiv r/R$ only and are given in an implicit analytical form by the
following equations: 
\begin{eqnarray}
\zeta^5&=&\paren{\frac{7}{6}V}^{-2}
\brace{\frac{7}{17}(5-3V)}^{-232/99}
\brace{\frac{7}{3}(4V-3)}^{5/11},\\
Z&=&\frac{2}{3}V^2(1-V)(4V-3)^{-1},\\
G&=& \frac{1}{49}(1-V)^{-3}\brace{\frac{7}{3}(4V-3)}^{9/11}
\brace{\frac{7}{17}(5-3V)}^{116/33}.
\end{eqnarray}
The pressure ration $p/p_2$ tends to a constant as $\zeta \to 0$, 
while $\beta/\beta_2 \propto \zeta$, 
$\rho/\rho_2 \propto \zeta^{9}$ in this limit.

%%%%%%%%%%%%%%%%%%%%%%%%%%%%%%%%%%%%%%%%%%%%%%%%%%%%%%%%%%%%%%%
\section{Numerical Results}
\label{numerical}
\subsection{Initial Conditions}
We consider an initial uniform spherical fireball surrounded by a
uniform cold ISM. The initial conditions are determined by four
parameters: the total energy $E$, the baryonic mass $M_0(=E/\eta)$ and
the radius $R_0$ of the fireball and the ISM density $\rho_1$.  A
``cold'' ISM means that its pressure is negligible compared to the
pressure behind the shocks throughout the whole evolutions. For
convenience, we set the initial time as $R_0$ rather than zero.  We
have chosen two sets of initial conditions to represent the the RRS
and the NRS cases: $E=10^{52}$erg, $\rho_1=1$proton $\mbox{cm}^{-3}$,
$\eta=50$ and $R_0=3\times 10^{10}$cm for the NRS case ($\xi=43$), and
$E=10^{52}$erg, $\rho_1=1$proton $\mbox{cm}^{-3}$, $\eta=10^4$ and
$R_0=4.3\times 10^{9}$cm for the RRS case ($\xi=0.1$).

We assume that the fluid is described by a constant adiabatic index
$\hat{\gamma}=4/3$ although in reality this is not alway true. The
shell's matter may cool during the coasting phase. However at the end
of the coasting stage, the reverse shocks heat the fireball shell, and
the ISM shocked by the relativistic forward shock carries most of the
energy $E$. The main part of the system is subject to relativistic
particles again. The forward shock is decelerated as $\Gamma \propto
R^{-3/2}$ and it becomes Newtonian at $l$.  After that the scaling
laws in the Newtonian regime depends on the adiabatic constant. But
even then the shocked electrons will remain relativistic for a long time
after this transition and the adiabatic index will remain around $4/3$.
%%%%%%%
\subsection{Numerical Results}
\begin{figure}[b!] % fig 2
\centerline{\epsfig{file=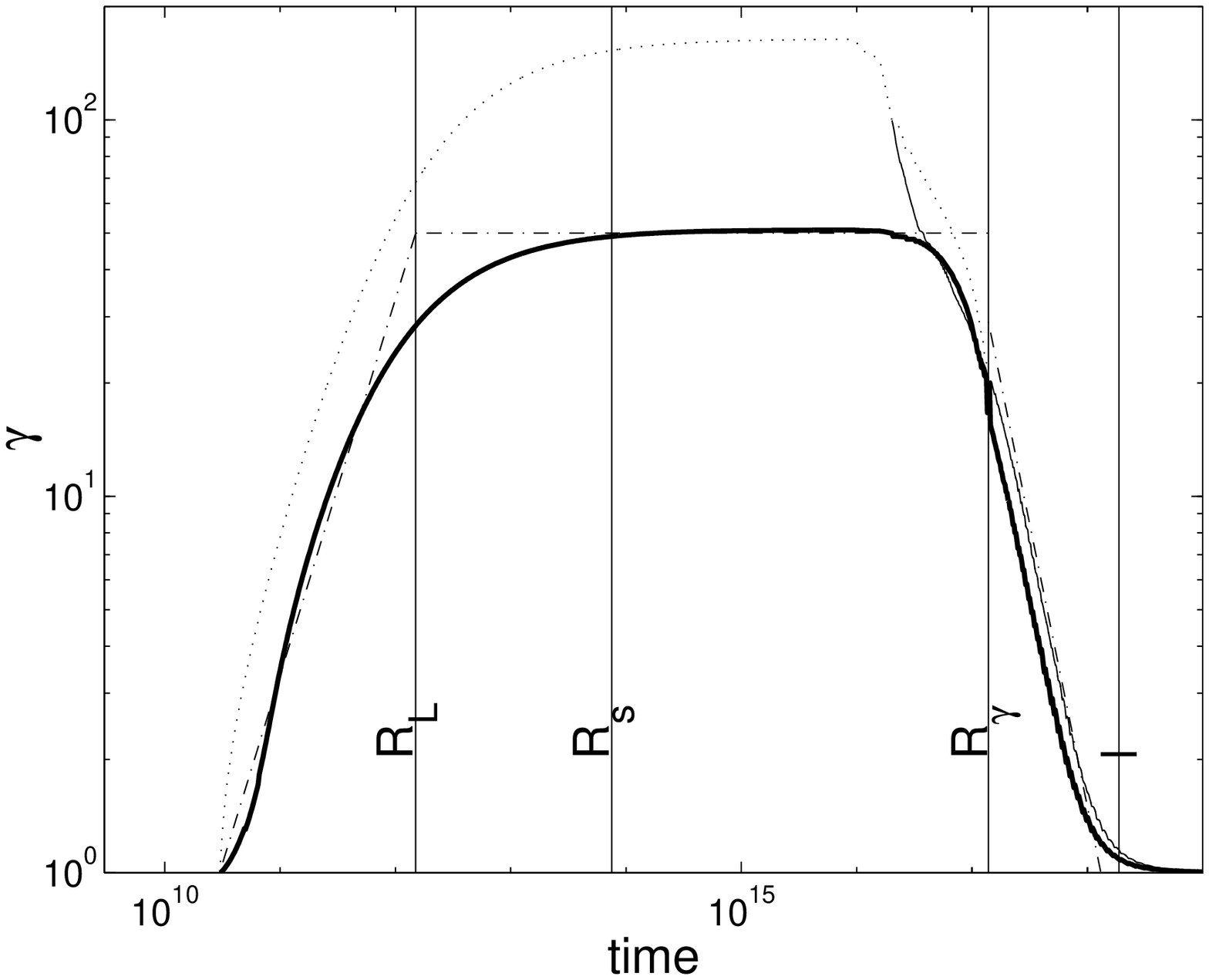,width=3.5in}}
\vspace{10pt}
\caption{$\gamma$ vs Time,  
for the NRS case ($\xi=43$, $E=10^{52}$ erg, $\eta=50$
$R_0=3\times10^{10}$cm).  The average value (thick solid
line), the value just behind the forward shock (thin solid line), the
maximal value (dotted line) and the analytic estimate (dashed dotted
line).  }
\label{fig:glob_n}
\end{figure}

\begin{figure}[b!] % fig 3
\centerline{\epsfig{file=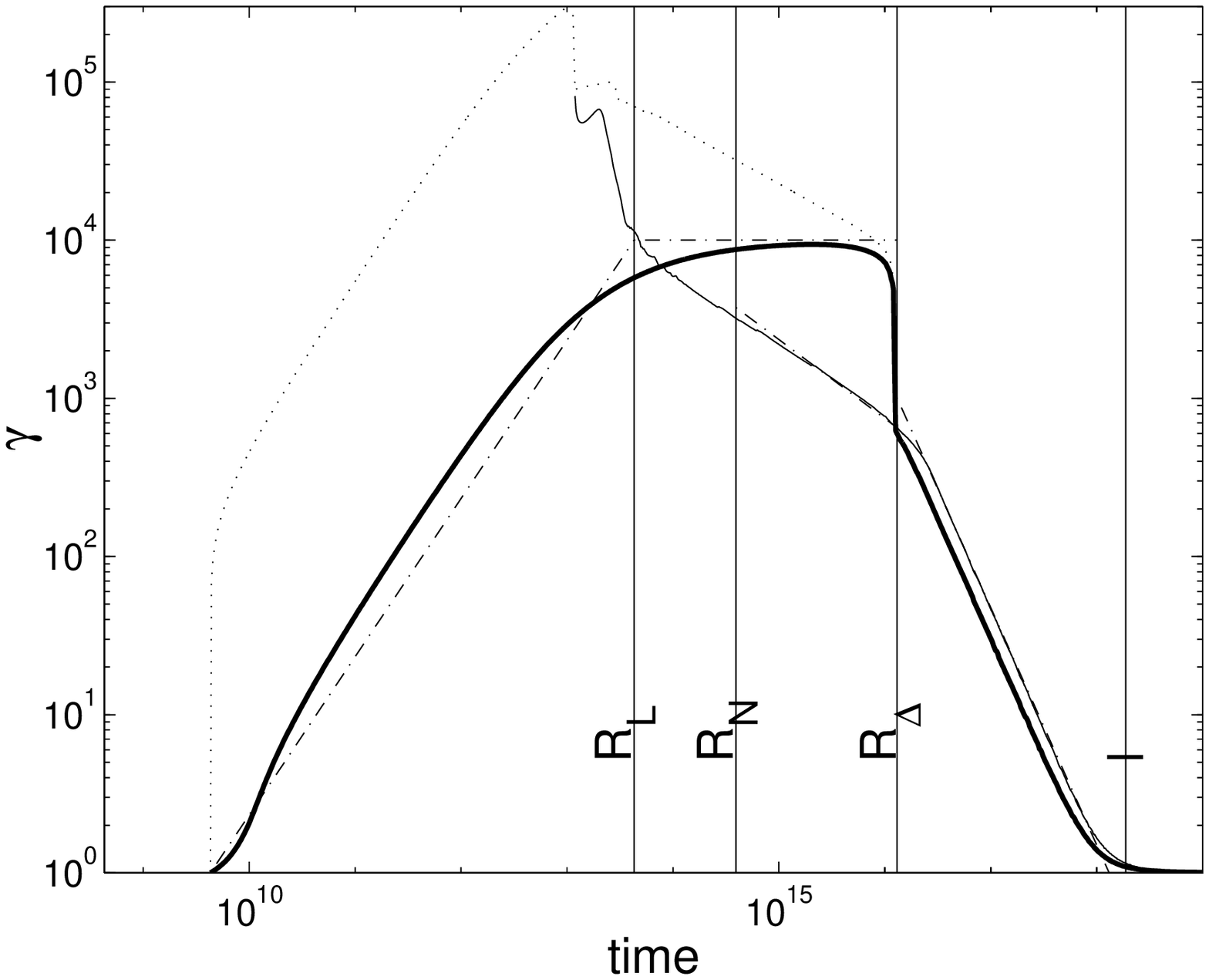,width=3.5in}}
\vspace{10pt}
\caption{$\gamma$ vs time for   
the RRS case ($\xi=0.1$, $E=10^{52}$ erg, $\eta=10^4$ 
$R_0=4.3\times10^{9}$ cm). The average value (thick solid
line), the value just behind the forward shock (thin solid line), the
maximal value (dotted line) and the analytic estimate (dashed dotted
line).  }
\label{fig:glob_r}
\end{figure}

Figures \ref{fig:glob_n},\ref{fig:glob_r} depict the evolutions of the
Lorentz factors for the NRS and the RRS cases, respectively.  One
clearly sees the initial acceleration phase in which the Lorentz
factor increases linearly with time. This is followed by a coasting
phase in which the Lorentz factor is a constant: $\eta$. This stage
ends when the effect of ISM becomes significant and most of the
kinetic energy is dissipated at $R_\gamma$ or $R_\Delta$. Then a self
similar phase begins in which the Lorentz factor of the forward shock
decreases like $\Gamma \propto R^{-3/2}$.  At $l$ the solution becomes
non relativistic and it turns into the Sedov-Taylor solution. A
difference appears between the NRS and the RRS fireballs only in the
energy transfer phase. As expected a sharp decrease in $\gamma$ is
seen in the RRS case. A more gradual transition  is seen in
the NRS case.

In the following sections we compare the numerical results with the 
analytic estimates. We examine the validity of the estimates of 
$R_L$, $R_s$, $R_\Delta$, $R_\gamma$, $R_N$ and $l$ as indicators of 
transition scales. 

%%%%%%%
\subsubsection{Free Acceleration Stage}
\begin{figure}[b!] % fig 4
\centerline{\epsfig{file=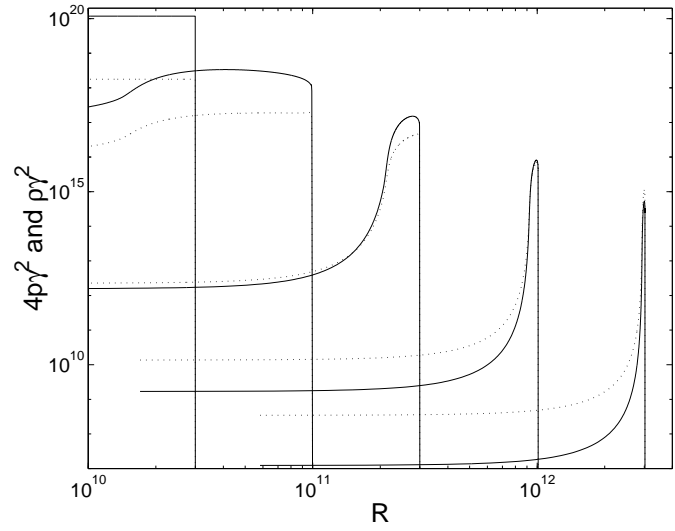,width=3.5in}}
\vspace{10pt}
\caption{Profiles of  the internal 
energy density $4p\gamma^2$ (solid line) and the mass density
$\rho\gamma^2$ (dotted line) in the frame of ISM during the initial
acceleration stage (internal energy dominated stage to matter
dominated stage).  The initial parameters are similar to those in
figure \ref{fig:glob_n}) }
\label{fig:pulse}
\end{figure}

\begin{figure}[b!] % fig 5
\centerline{\epsfig{file=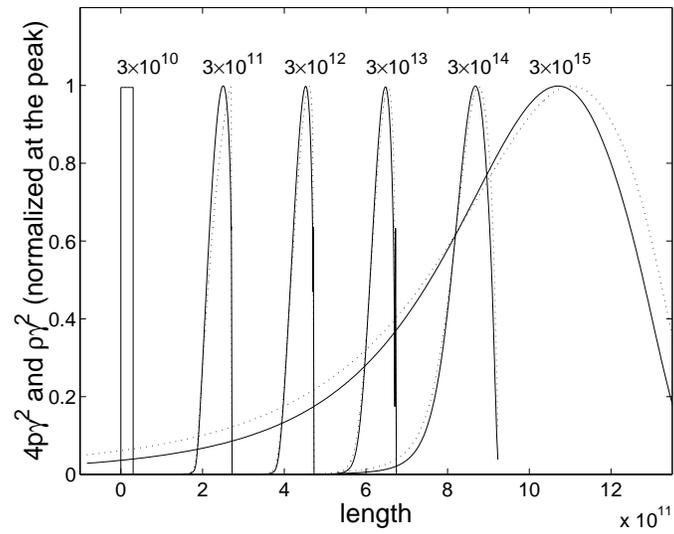,width=3.5in}}
\vspace{10pt}
\caption{The shell structures at different times
in the acceleration and coasting stage.  The internal energy density
$4p\gamma^2$ (solid line) and the mass density $\rho\gamma^2$ (dotted
line) normalized at the peaks. The initial parameters are similar to
those in figure \ref{fig:glob_n}) The initial time is set as $R_0
(=3\times 10^{10})$.}
\label{fig:pulse_comp}
\end{figure}

A highly relativistic shell is formed after a short acceleration phase
(see figure \ref{fig:pulse}). The width of the shell is 
constant in this acceleration stage (figure \ref{fig:pulse_comp}).
The density and the pressure  peak at the same position in the local 
fluid frame. In the observer frame, the outer part of the shell has
a higher Lorentz factor and the density peak is wider than the pressure
peak. The density peak (dotted line) moves ahead of the pressure peak 
(solid line) in the observer frame (figure \ref{fig:pulse_comp}).

\begin{figure}[b!] % fig 6
\centerline{\epsfig{file=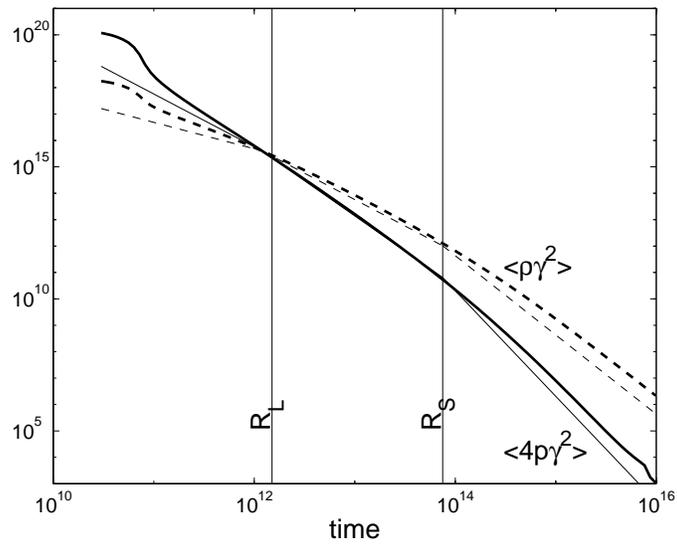,width=3.5in}}
\vspace{10pt}
\caption{  The average
internal energy density $4p\gamma^2$ (thick solid line), the average
mass density $\rho\gamma^2$ (thick dashed line) and an analytical
estimate of the internal energy density (thin solid line) and the mass
density (thin dashed line). The latter are normalized at the cross
point of the numerical results.  The initial parameters are the same
as in figure
\ref{fig:glob_n}.  }
\label{fig:pulse_max}
\end{figure}

The average Lorentz factor of the each fluid element in the shell
increase as the radius of the shell increases, $\gamma \sim
r/\Delta$. However, the Lorentz factor of the outermost layers is well
above the average. This is results from the initial sharp edges of the
fireball.  The initial acceleration of the outermost layers depends on
the steepness of the initial pressure distribution at the edge of the
fireball. The initial step function distribution that we have chosen
leads to large acceleration of the outermost layers.  We can regard
the thin region at the boundary of the fireball as expanding in the
free expansion velocity, independent of the fireball thickness
$\Delta$. However, this fast layer is thin and its mass is negligible.
Except of the evolution of the maximal Lorentz factor (at the
outermost layers) the evolution of the bulk of the fireball is not
effected by the choice of the initial steepness of the boundary. The
initial conditions are washed out later, when the interaction with the
ISM is significant, and even the maximal Lorentz factor is then
independent of the initial conditions.  Therefore we do not discuss
this maximal value, but instead we consider the average Lorentz factor
over the ``uniform shell''. We define the average value as:
\begin{equation}
\angle{f}\equiv\int f m r^2 dr /\int m r^2 dr,
\label{eq:ave}
\end{equation}
where $m \equiv \gamma \brace{\rho+(3+\beta^2)p}$ 
is the effective mass density in
the observer frame and the integrals are defined from the origin to a
radius at which the Lorentz factor takes the maximal value.  The
evolution of the average Lorentz factor $\angle{\gamma}$ for the NRS
and the RRS cases is shown in figure \ref{fig:glob_n} and
\ref{fig:glob_r} (thick line) respectively.  The Lorentz factors
increase linearly with time. The average internal energy and the mass
density in the observer frame are shown in figure
\ref{fig:pulse_max}. $R_L$ is a good indicator to the transition from
the acceleration stage to the coasting stage.  The mass density equals
the internal energy density at $\sim R_L$ (see figure
\ref{fig:pulse_max}).

%%%%%%%
\subsubsection{Coasting Stage}
After the fireball uses up all the internal energy at $R_L$, it coasts
with a Lorentz factor $\eta$ (see figures \ref{fig:glob_n} and
\ref{fig:glob_r}). In the RRS case the shell has a frozen radial
profile through this stage, while in the NRS case the frozen pulse
approximation breaks down at $R_s$ and the shell begins to expand
before the ISM has most of the system energy at $R_\gamma$ (see figure
\ref{fig:pulse_comp}).  We can see the transition of the scaling laws
of the density and the pressure at $R_s$ in figure
\ref{fig:pulse_max}.

%%%%%%%
\subsubsection{Energy Transfer Stage: The RRS Case}
\begin{figure}[b!] % fig 7
\centerline{\epsfig{file=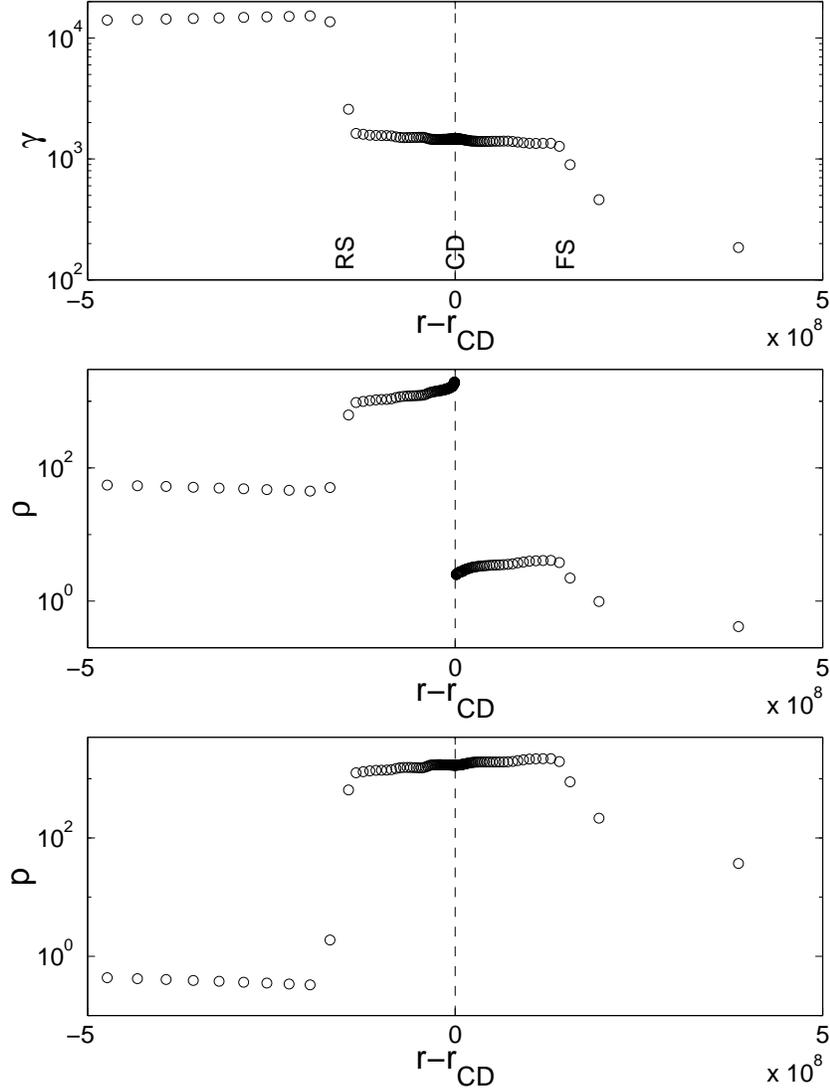,width=5in}}
\vspace{10pt}
\caption{ : 
The Lorentz factor $\gamma$, density $\rho$ and pressure $p$ in the
deceleration stage by the reverse shock ($t=3\times 10^{15}$) for an
RRS solution. The x axis is the distance from the contact
discontinuity (dashed line). RS and FS indicate the positions of the
reverse shock and the forward shock, respectively.  The initial
parameters are the same as in figure \ref{fig:glob_r}.  }
\label{fig:rs_pro}
\end{figure}

In the RRS case the reverse shock becomes relativistic at $R_N$ before
it crosses the shell at $R_\Delta$. From this moment onwards the
Lorentz factor of the coasting shell is reduced considerably after the 
passage of the  reverse shock. Figure \ref{fig:rs_pro} shows the
deceleration by a  relativistic reverse shock. There are four regions
in the figure: the ISM, the shocked ISM, the shocked shell and the
unshocked shell, which are separated by the forward shock (FS), the
contact discontinuity (CD) and the reverse shock (RS).

\begin{figure}[b!] % fig 8
\centerline{\epsfig{file=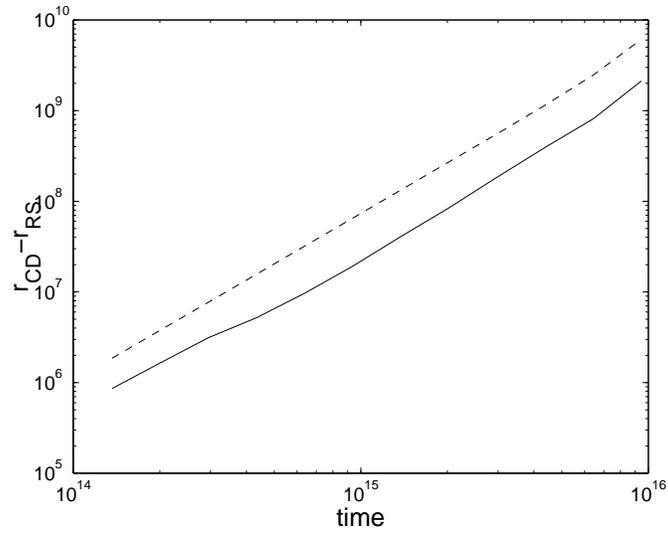,width=3.5in}}
\vspace{10pt}
\caption{ The  
distance between the contact discontinuity and the reverse shock as a
function of time: $r_{CD}-r_{RS}$ (solid line), $t/2\gamma_4\sqrt{f}$
(dashed line) for a RRS solution.  The initial parameters are the same
as in figure \ref{fig:glob_r}.  }
\label{fig:rshock}
\end{figure}

Using the shocks' jump conditions and the equality of the pressure and
the velocity at the contact discontinuity, we can estimate the Lorentz
factor of the shocked region, $\gamma_{CD}$, and the Lorentz factor of
the reverse shock, $\Gamma_{RS}$, for a planner geometry (Sari \& Piran,
1995),
\begin{equation}
\gamma_{CD}= f^{1/4}\sqrt{\gamma_4/2}, \ \ 
\Gamma_{RS}=\gamma_{CD}/\sqrt{2},
\label{eq:gamma_2}
\end{equation}
where $f\equiv \rho_4/\rho_1$. The quantities just ahead of (or just
behind) the reverse shock are denoted by the subscript 4 (or 3).  For
spherical geometry, the pressure is not constant in the shocked region
bounded by the two shocks. The ratio $f$ in the above equations should
be replaced with $(p_2/p_3) f$.  However, in our simulations,
$p_2/p_3$ is a factor of a few at most. Then, equation
(\ref{eq:gamma_2}) can roughly explain the numerical result. The time
it takes for the reverse shock to cross a distance $dx$ in the shell
material is $dt=dx \gamma_4
\sqrt{f}/2$. As the reverse shock compresses the shell material, the
width $dx$ becomes $dx/2$ after the shock passes through it. Then, the 
distance between the contact discontinuity and the reverse shock is 
$t/2\gamma_4\sqrt{f}$ \ \  ($\propto t^2$). Figure \ref{fig:rshock} 
shows the numerical result.

\begin{figure}[b!] % fig 9
\centerline{\epsfig{file=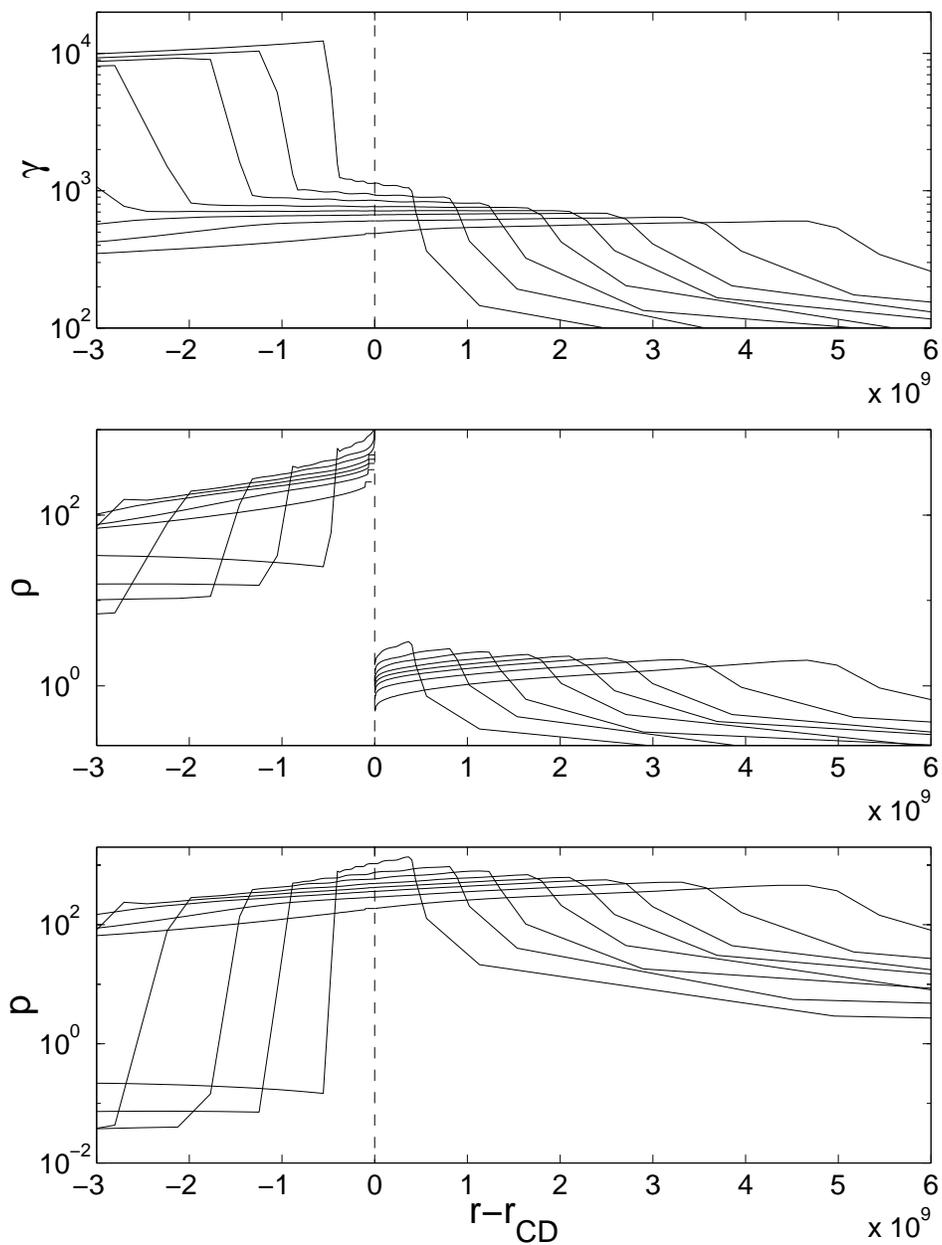,width=5in}}
\vspace{10pt}
\caption{ The transition into the Relativistic
self similar solution in a RRS solution. Profiles of $\gamma$, $\rho$
and $p$ at different times 
($t=0.41 R_\Delta$, $0.60 R_\Delta$, $0.73 R_\Delta$, $0.85 R_\Delta$,
$0.93 R_\Delta$, $R_\Delta$, $1.1 R_\Delta$ and $1.2 R_\Delta$) 
The x axis is the distance from the 
contact discontinuity (dashed line).  }
\label{fig:into_bm}
\end{figure}

The relativistic reverse shock passes through the shell as $t^2$
and it decelerates the coasting shell drastically. After this drastic 
deceleration, the shocked shell slows down as $t^{-1/2}$ 
(figure \ref{fig:glob_r}) due to the pressure difference between $p_2$
and $p_3$. At $R_{\Delta}$ when the reverse shock crosses the shell, 
$\gamma_{CD}$ reaches effectively the value of $\gamma_2$ expected
from the relativistic self similar solution and the profile of the 
shocked ISM region begins to approach the self similar one. 
The transition into the self similar solution is shown 
in figure \ref{fig:into_bm}. 

At the beginning of the self similar
deceleration stage, the density of the shocked shell is 
much larger than that of the shocked ISM. There is a large gap of 
the density at the contact discontinuity. However, the density
perturbation, as we discuss in section \ref{perturbation}, does 
not effect $\gamma$ and $p$, and it does not propagate in the local
fluid frame. Then, as the blast wave expands, it leaves the gap, and
the ratio between $\rho_2$ and the density of the shocked shell damps.

%%%%%%%
\subsubsection{Energy Transfer Stage: The NRS Case}
\begin{figure}[b!] % fig 10
\centerline{\epsfig{file=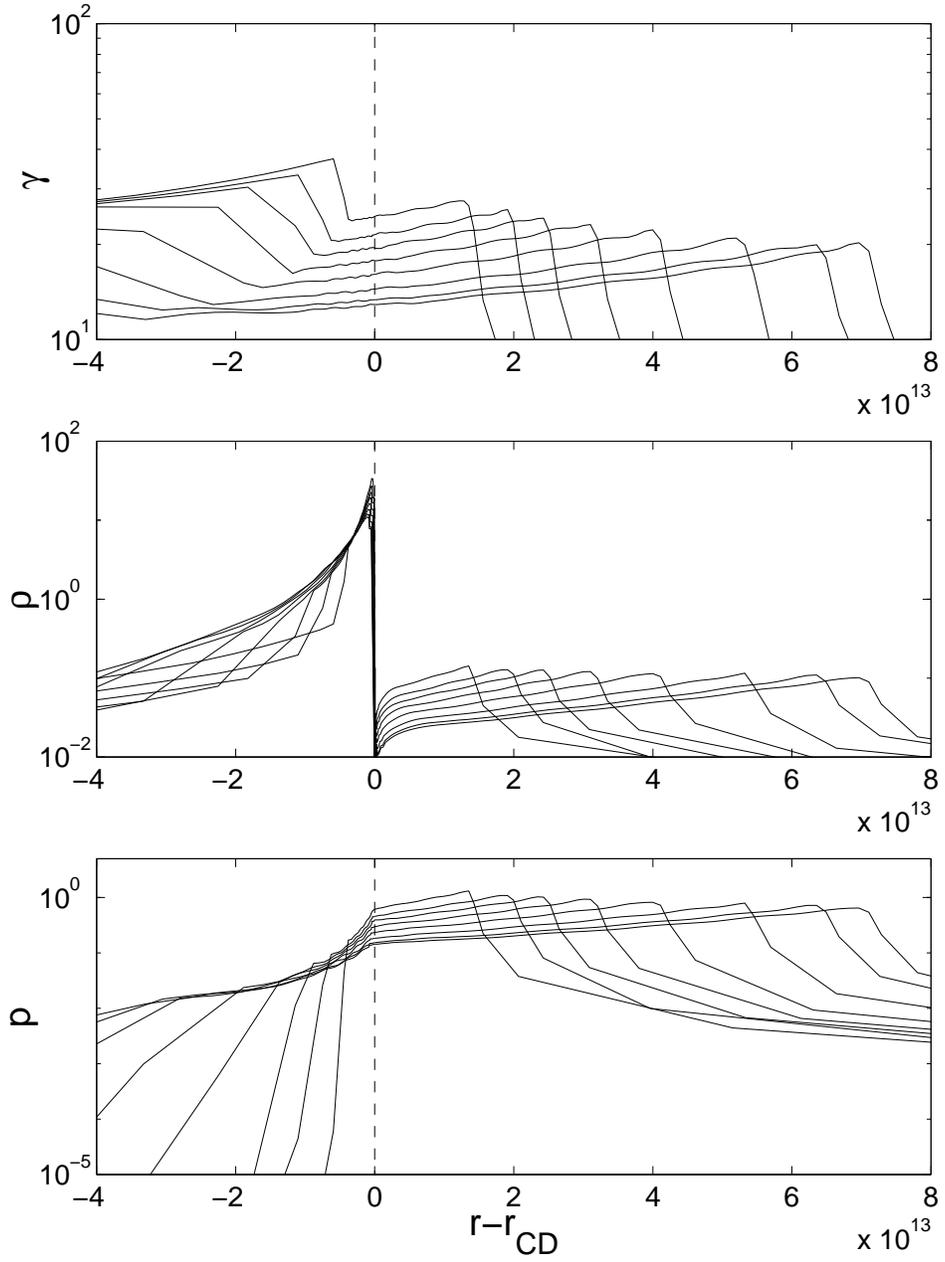,width=5in}}
\vspace{10pt}
\caption{ The transition into the Relativistic
self similar solution for a NRS solution.  Profiles of $\gamma$,
$\rho$ and $p$ at different times ($t=0.70R_\gamma$, $0.77R_\gamma$, 
$0.82 R_\gamma$, $0.87 R_\gamma$, $0.92 R_\gamma$,
$0.98 R_\gamma$, $R_\gamma$, and $1.1 R_\gamma$ ) 
The x  axis is the distance from the
contact discontinuity (dashed line).  }
\label{fig:into_bm_n}
\end{figure}

In the NRS case ($\xi > 1$), the shell begins to spread at $R_s$ and
the value of $\xi$ decreases.  At $R_\gamma$, a coincidence
$R_\Delta=R_\gamma=R_N$ happens. The shocked ISM becomes the main 
component of the system and the profile approaches the Blandford-McKee
solution. At this stage the reverse shock is 
just mildly relativistic, the shell's Lorentz factor changes
by a factor of order unity (see figure \ref{fig:into_bm_n}).

%%%%%%%
\subsubsection{Relativistic Self Similar Deceleration Stage}
\begin{figure}[b!] % fig 11
\centerline{\epsfig{file=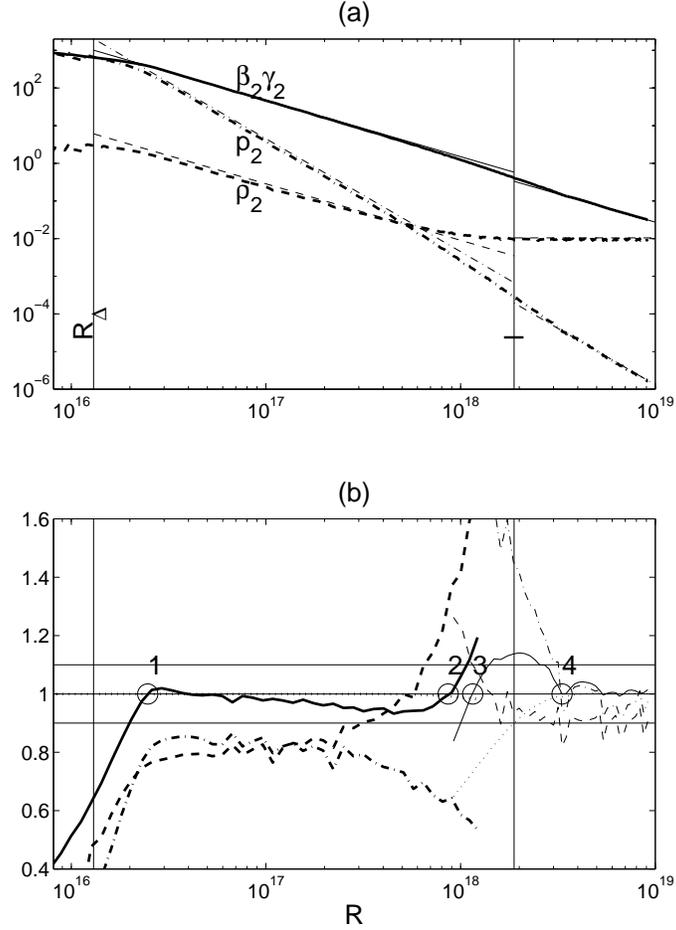,width=3.5in}}
\vspace{10pt}
\caption{The deceleration stage. (a)$\beta_2 \gamma_2$ (solid lines), 
density (dashed lines) and pressure (dashed dotted lines) just behind
the shock vs the shock radius.  The numerical results are thick lines and
the relativistic and the Newtonian self similar solution are thin
lines. We plot $\gamma_2$ ($\beta_2$) rather than $\beta_2\gamma_2$
for the relativistic self similar solution (the Newtonian self similar
solution).  The initial parameters are the same as in figure
\ref{fig:glob_r}.  }
(b) The ratio between the numerical results and the analytic
estimates. $\gamma_2$ or $\beta_2$ (solid line), density (dashed
line) and pressure (dashed dotted line) just behind shock and the
shock radius (dotted line).
The numerical results are compared with the relativistic (thick line) 
or the Newtonian (thin line) self similar solution. $\gamma_2$
($\beta_2$) is compared with the analytic estimate in the relativistic
regime (the Newtonian regime). The circles 1,2,3 and 4
indicating the cross points are at 
$R=1.9 R_\Delta, 
0.46 l, 0.61 l$ and $1.8 l$ respectively. 
\label{fig:bm_stage}
\end{figure}

In the deceleration stage, the hydrodynamic profiles of the 
shocked ISM depends on only $E$ and $\rho_1$. The numerical values 
of the Lorentz factor $\gamma_2$, the density $\rho_2$ 
and the pressure $p_2$ just behind the shock are compared with 
the self similar solution in figure \ref{fig:bm_stage}. 
The numerical result of Lorentz factor is consistent with the 
Blandford-McKee self similar solution within a few $\%$ difference
in the relativistic regime (see figure \ref{fig:bm_stage} (b)).  
Though the density and the pressure peaks are narrower than the 
velocity peak (see equation (\ref{eq:bm})) and 
the numerical errors are larger, the density and the pressure 
agree with the analytic estimates within $\sim 20\%$.
In the RRS case, $\gamma_2$ begins to satisfy the   
relativistic self similar scaling 
at $R=1.9 R_\Delta$ (circle 1 in figure \ref{fig:bm_stage} 
(b)). The self similar solution and the equation (\ref{eq:gammacd}) with
a factor $1/2\pi^{1/4}$ which we neglected for simplicity, 
give the analytic estimate $1.5 R_\Delta$. In the NRS case, 
$\gamma_2$ reaches within a $10\%$ error line at $R=2.4 R_\gamma$.

\begin{figure}[b!] % fig 12
\centerline{\epsfig{file=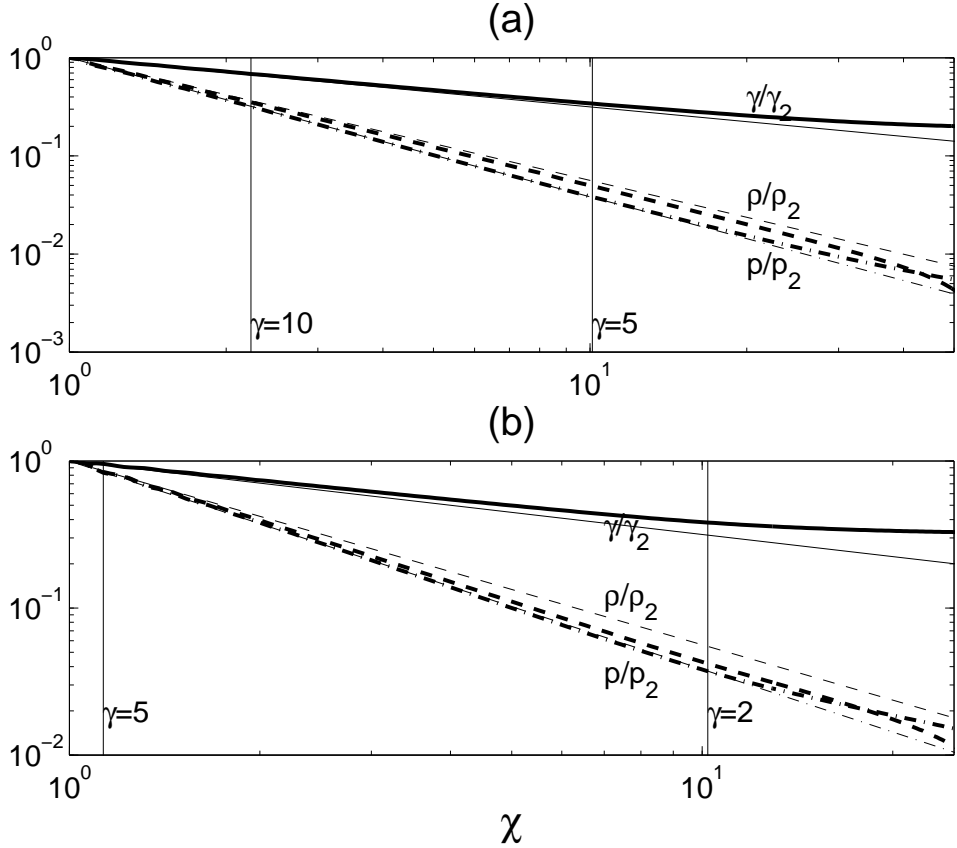,width=5in}}
\vspace{10pt}
\caption{The Lorentz factor (solid lines), the density (dashed lines)  
and the pressure (dashed dotted lines) as a function of $\chi$ during
the relativistic deceleration stage at different times: (a)$t=2\times
10^{17}$ ($\gamma_2\sim15$) (b)$t=4\times 10^{17}$ ($\gamma_2\sim8$):
The numerical results are thick lines and the analytical results are
thin lines.  $\gamma$, $\rho$ and $p$ are normalized with the
numerical results $\gamma_2$,$\rho_2$ and $p_2$.  The vertical lines
indicate the positions where the fluid elements have the values of the
Lorentz factor $\gamma=2, 5$ or $10$. 
The initial parameters are the same as in figure
\ref{fig:glob_r}.  }
\label{fig:bm_chi_comb2}
\end{figure}

Flow profiles are plotted as a function of the 
similarity variable $\chi$ in figure \ref{fig:bm_chi_comb2}.
In the top panel of the figure, the value of the Lorentz factor 
just behind the forward shock is $\sim 15$, and the value drops
to $\sim 5$ in the inner region. The numerical results agree well 
with the self similar solution.
In the bottom panel of the figure, the value of the Lorentz factor 
just behind the forward shock is $\sim 8$, and the value drops
to $\sim 2$ at the inner region where the numerical results 
deviate from from the self similar solution.
The relativistic self similar solution is derived with an assumption 
that each fluid element is highly relativistic, but it is a good 
approximation still at $\gamma \sim 5$.

%%%%%%%
\subsubsection{Transition to the Sedov-Taylor Solution}
The Lorentz factor of the forward shock decreases and it becomes
non-relativistic around $l (\sim 1.9 \times 10^{18}$ in this
case). The scaling laws of the velocity $\beta_2$, the density
$\rho_2$ and the pressure $p_2$ also gradually shift from the
Blandford-McKee solution to the Sedov-Taylor solution around $l$ (see
figure \ref{fig:bm_stage}). The circles 2 ($R_{c2}\equiv 0.46 l$) and
3 ($R_{c3} \equiv 0.61 l$) in figure \ref{fig:bm_stage} (b) give a
rough estimate of the radii where the relativistic self similar
solution becomes invalid and the Newtonian self similar solution
becomes valid ($\gamma_2 \sim 1.9$ at circle 2, $\beta_2 \sim 0.70$ at
circle 3). The relation between $\beta_2$ and $R$ in the Newtonian
self similar solution is already valid at circle 3, but the shock
radius $R$ is still proportional to time $t$ at this stage and it is
smaller than the radius expected by the Newtonian self similar
solution. The relation between $\beta_2$ and $t$ in the Newtonian self
similar solution becomes valid at circle 4 ($R_{c4}\equiv 1.8 l$)
where $\beta_2=0.14$.

\begin{figure}[b!] % fig 13
\centerline{\epsfig{file=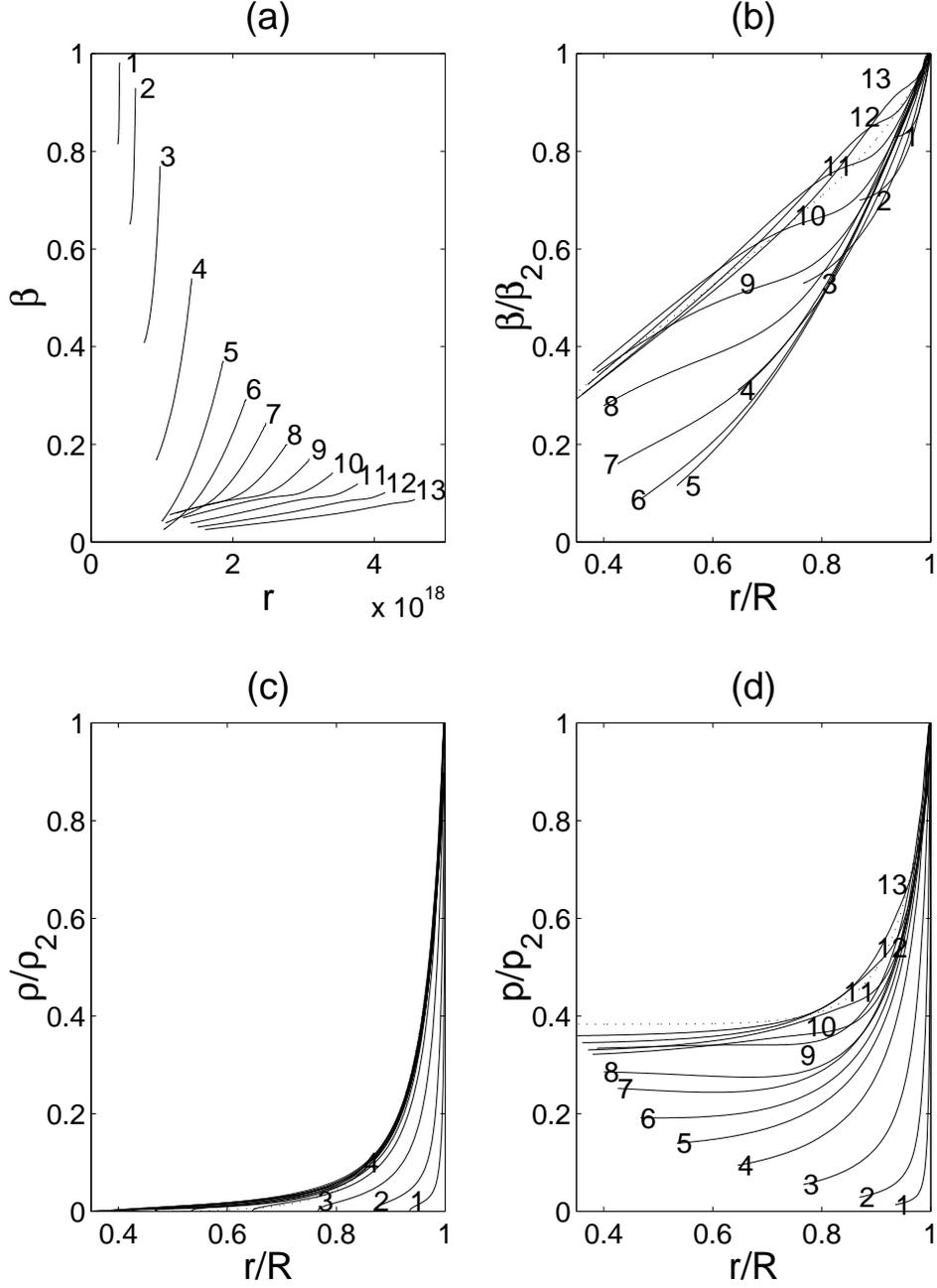,width=5in}}
\vspace{10pt}
\caption{The transition of the flow profile of the shocked ISM from 
the relativistic Blandford-McKee solution  to the Newtonian Sedov-Taylor
solution: 
(a) $\beta$ vs $r$ (b) $b/b_2$ vs $r/R$ (c) $\rho/\rho_2$ vs $r/R$
and (d) $p/p_2$ vs $r/R$.
The dotted lines shows the Newtonian self similar solution.
The number denote the  shock radius $R=$:
1)  $0.21 l$, 
2)  $0.33 l$,
3)  $0.52 l$,
4)  $0.76 l$,
5)  $1.0 l $,
6)  $1.2 l $,
7)  $1.3 l $,
8)  $1.5 l $,
9)  $1.6 l $,
10) $1.8 l $,
11) $2.0 l $,
12) $2.2 l $,
13) $2.4 l $.
The initial parameters are the same as in figure \ref{fig:glob_r}.
}
\label{fig:all_zeta}
\end{figure}

The transition of the profiles in the shocked ISM region, from the
relativistic stage to the Newtonian stage, is shown in figure
\ref{fig:all_zeta}. The jump condition for a strong shock gives a
simple relation between the ISM density $\rho_1$ and the shocked
density $\rho_2^{\prime} (=\rho_2\gamma_2)$ measured in the observer
frame, $\rho_2^{\prime}=\gamma_2(4\gamma_2+3)\rho_1$.  If ISM have
been swept up at a radius $R$, the thickness of the blast wave is
approximately $R/\gamma_2(4\gamma_2+3)$ $\propto R^4$ for the
relativistic stage. The thickness is $\propto R$ for the Newtonian
stage. As the blast wave expands, it becomes broader (figure
\ref{fig:all_zeta}a). The inner part of the distribution begins 
to approach the Newtonian self similar one Around $R_{c3}$ (figure
\ref{fig:all_zeta}b).  The density profile  approaches  the one
expected by the Newtonian self similar solution at $R_{c3}$ (see
figure \ref{fig:all_zeta}c). The pressure profile approaches the
Newtonian one more gradually and it still evolves after $R_{c3}$
(figure \ref{fig:all_zeta}d). The velocity and pressure profiles are
consistent with the analytic estimates within $10 \%$ level at
$R_{c4}$.

%%%%%%%
\section{Evolution of Perturbations to Blandford-McKee Solution}
\label{perturbation}
The simulations presented in  the previous sections have shown that the
hydrodynamical profiles of the shocked ISM approaches the
relativistic self similar Blandford-McKee solution at the end of the
energy transfer phase. It implies that the self similar solution is
attractive and stable. In this section we consider the evolution 
of spherical perturbations to the self similar profile.

For an infinite uniform fluid there are two types of perturbations 
in the linear theory. One type is a sound wave propagating with 
the sound velocity relative to the fluid. The other is an 
entropy-vortex wave moving with the fluid. It is an entropy 
perturbation with no change of the pressure and it includes a
spherical density perturbation.

\begin{figure}[b!] % fig 14
\centerline{\epsfig{file=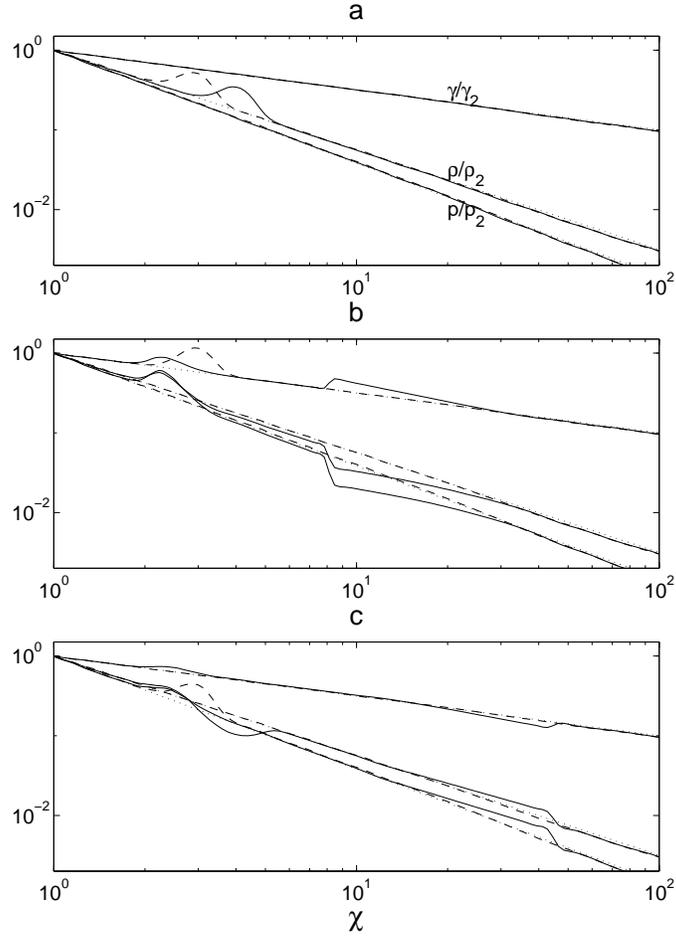,width=3.5in}}
\vspace{10pt}
\caption{Stability of the Relativistic self similar
solution.  Perturbation of $\rho$ - (a), $\gamma$ - (b) and $p$ - (c).
Shown are the unperturbed self similar solution (dotted lines), the
initial perturbed profiles (dashed lines) and the future evolution
(solid lines) }
\label{fig:stab}
\end{figure}

As an unperturbed initial configuration we take a self similar blast
wave with a total energy $E=10^{52}$ erg in a uniform ISM of the
density 1 proton $/cm^3$, at the moment where the Lorentz factor
behind the shock is $\gamma=4000$ ($t \sim 5.3\times 10^{15}$). We
then add a Gaussian perturbation $\delta q/q=\delta \
exp\bra{-(\chi-\chi_0)^2/\Delta \chi^2}$, with $\delta=1, \chi_0=3$
and $\Delta \chi=0.5$ to one of the hydrodynamical variables: the
Lorentz factor, the density or the pressure.  $q$ stands for these
variables and $\delta q$ for its perturbation.  Figures \ref{fig:stab}
(a), (b) and (c) depict the flow profiles as functions of $\chi$ for
the perturbation of $\rho$, $\gamma$ and $p$, respectively. The dashed
lines are the initial profile (self similar solution + perturbation),
and the solid lines are at $\delta t=4.3\times 10^{14}$sec later.

The density perturbation (figure \ref{fig:stab} (a)) does not effect
other quantities ($\gamma$ and $p$) and it does not propagate in the
local fluid frame. We use the self similar solution to estimate the
evolution of $\chi_e$ of the fluid element in the perturbed region (or
any other fluid element). Taking the derivative of $\chi_e$ along its
line of flow we get $d\chi_e/dt=4(\chi_e+1)/t$ so that a fluid element
which had been at $\chi_0$ at a time $t_0$, will be at $\chi_e(t) \sim
\chi_0(t/t_0)^4$. The fluid element which had a density $\rho_0$ at
$t_0$, will have a density $\rho_e(t)=\rho_0(t/t_0)^{-13/2}$. The
ratio between the density perturbation $\delta \rho_e(t,r(t))$ and the 
unperturbed value $\rho_e(t, r(t))$ is constant in time, but the
perturbation will depart from the forward shock as $\chi_e \propto t^4$  
and the perturbation becomes less important as 
$\rho_e/\rho_2 \propto t^{-5}$.  Our numerical results agree well
with these scalings. 
 
A velocity perturbation (figure \ref{fig:stab} (b)) induces pressure
and density perturbations. The pulse of the coupled perturbations is
decomposed after a short time into two pulses: An outgoing compression
pulse (outgoing motion corresponds to a decreasing $\chi$) and an
ingoing rarefaction pulse (a rarefaction wave and a shock wave).
These waves propagate with the speed of sound in the local fluid frame
$\gamma^{\prime}_s=\sqrt{4p_e/3\rho_e}=2\gamma_2^{1/2}\chi^{-1/12}/3$.
The speed of the outgoing and the ingoing sound wave in the observer
frame are $\gamma_s=4\gamma_2^{3/2}\chi^{-7/12}/3$ and
$3\gamma_2^{1/2}\chi^{-5/12}/4$, respectively.  Assuming that the
initial position $\chi_0$ of the perturbation at a time $t_0$ is not
too far from the forward shock, $\gamma_s \gg \Gamma$, we get the
position of the outgoing pulse $\chi=\chi_0(t/t_0)^{-4}$. The pulse
reaches the forward shock at $t=\chi_0^{1/4}t_0$ and it boosts the
forward shock. The position of the ingoing pulse cannot be
expressed by a simple analytic formula, but it departs from the
forward shock much faster than a density perturbation. After the
out-going pulse boosts the forward shock, the flow profile is almost
the self similar one.

A pressure perturbation (figure \ref{fig:stab} (c)) also induces
perturbations in $\gamma$ and $\rho$.  The perturbations
consist of three components, two propagating component and a standing
density perturbation.  After an outgoing compression pulse and an 
ingoing shock propagate, a density perturbation still stays at the
position where the initial pressure perturbation was (in the local 
fluid frame).  The ingoing shock leaves the forward shock quickly. 
The density perturbation departs from the forward shock as discussed 
earlier. The outgoing pulse is basically the same one appearing in 
the case of an initial velocity perturbation. The pulse boosts the 
forward shock and at this stage the profile is almost the self similar
one.

%%%%%%%
\section{Conclusions}
\label{conclusions}

We have explored numerically the evolution of a relativistic fireball
interacting with a uniform ISM, through the stages of initial
acceleration, coasting, energy transfer to the ISM and the
deceleration. These calculations begin when the fireball is at rest.
They follow the acceleration to a relativistic velocity and the
subsequent slowing down to a velocity far below the speed of
light. These calculations span more than eight orders of magnitudes in
the size of the fireball.  The current analytic understanding of the
fireball evolution can explain well our numerical results. Initially,
the Lorentz factor increases linearly with the radius during the
initial free acceleration stage. At $R_L$ the fireball has transfered
all its initial radiation energy to kinetic energy and it coasts. Then the 
energy is transfered to the ISM. This takes place at $R_{\gamma}$ for
the NRS case, or at $R_{\Delta}$ for the RRS case. After that the
shocked ISM carries most of the initial energy of the fireball. The
profile of the shocked ISM is then described well by the relativistic
self similar Blandford-McKee solution. The forward shock decelerates
as $\Gamma
\propto R^{-3/2}$ and it becomes Newtonian at $l$. After that the
non-relativistic self similar Sedov-Taylor solution sets in. We have
shown that the relativistic self similar solution is an excellent
approximation, down to $\gamma \sim 5$ and a reasonable approximation
even down to $\gamma \sim 2$.  We have also examined the transition
into the relativistic self similar solution, and the transition from
this solution to the non-relativistic self similar solution.

With the spherical code, we have shown that the hydrodynamical 
profiles of the shocked ISM approaches to the relativistic 
self similar solution at the end of the coasting stage,
even though the initial conditions of the simulation do not have
self similar profiles. It implies that the relativistic self similar 
solution is attractive and stable for a spherical perturbation. We
have tracked the evolution of the spherical density, velocity and
pressure perturbations to the relativistic self similar solution
in order to verify the stability of this solution.

The Blandford-McKee solution is the basis of much of the GRB afterglow
theory. It is used to obtain an explicit expression for the radial
profile during the deceleration stage. We have shown that this model
is reasonable even down to $\gamma \sim 2$. It is also valid even in
the case of a radial inhomogeneity in the ISM. 
%A blast wave in an external density gradient $\rho_1
%\propto r^{-k}$ is another type of the Blandford-McKee solution.

Though the results presented in this paper are intended as a more
general explanation of the strong explosion problem rather than a
detailed fits of the GRB afterglow, comparisons of the numerically
predicted light curves based on this computations in detailed
realistic models with the recent and upcoming observations will enable
us to determine free parameters in the GRBs model that cannot be
calculated from first principles, such as the total energy of the
system, the surrounding density, the fraction of the energy that is
given by the shocks to electrons or to the magnetic field.

\ \ 

%%%%%%%%%%%%%%%%%%%%%%%%%%%%%%%%%%%%%%%%%%%%%%%%%%%%%%%%%%%%%%%%%%%%
S.K. gratefully acknowledges the support by the Golda Meir Postdoc 
fellowship. R.S. thanks The Clore Foundations for support. This work
was supported in part by a US-Israel BSF grant 95-328 and by a NASA 
grant NAG5-3516.
%%%%%%%%%%%%%%%%%%%%%%%%%%%%%%%%%%%%%%%%%%%%%%%%%%%%%%%%%%%%%%%%%%%%
\newpage 
\noindent {\bf References}\\
\noindent Blandford,R.D. \& McKee,C.F. 1976, 
          Phys. of Fluids, 19, 1130.\newline
Daigne, F. \& Mochkovitch, R. 1997, preprint.\newline
Goodman,J. 1986, ApJ, 308, L47.\newline
Kobayashi,S., Piran,T. \& Sari,R. 1997, ApJ, 490, 92.\newline
Landau,L.D. \& Lifshitz,E.M. 1987, Fluid Mechanics 2nd ed.
(Pergamon Press), Chap. X.\newline 
Katz, J., 1994, ApJ,  422, 248..\newline
M\'esz\'aros,P., Laguna,P. \& Rees,M.J. 1993, ApJ, 415, 181.\newline
M\'esz\'aros,P. \& Rees,M.J. 1992, ApJ, 397, 570.\newline
Narayan,R., Paczy\'nski,B. \& Piran,T. 1992, ApJ, 395, L83.\newline
Paczy\'nski,B. 1986, ApJ, 308, L51.\newline
Piran,T., Shemi,A. \& Narayan,R. 1993, MNRAS, 263, 861.\newline
Rees,M.J. \& M\'esz\'aros, P. 1994, ApJ, 430, L93.\newline
Sari,R. 1997, ApJ, 489, L37.\newline
Sari,R. \& Piran,T. 1995, ApJ, 455, L143.\newline
Sari,R. \& Piran,T. 1997, ApJ, 485, 270.\newline
Sedov,L.I. 1946, Prikl. Mat. i Mekh.,10, 241\newline
Shemi,A. \& Piran,T. 1990, ApJ, 365, L55.\newline
Taylor G.I. 1950, Proc. Roy. Soc. London, A201, 159.\newline
Von Neumann J. 1947, Los Alamos Sci. Lab. Tech. Series, Vol 7.\newline
\newpage
%%%%%%%%%%%%%%%%%%%%%%%%%%%%%%%%%%%%%%%%%%%%%%%%%%%%%%%%%%%%%%%
%%%%%%%%%%%%%%%%%%%%%%%%%%%%%%%%%%%%%%%%%%%%%%%%%%%%%%%%%%%%%%%
%%%%%%%%%%%%%%%%%%%%%%%%%%%%%%%%%%%%%%%%%%%%%%%%%%%%%%%
%%%%%%%%%%%%%%%%%%%%%%%%%%%%%%%%%%%%%%%%%%%%%%%%%%%%%%%
%%%%%%%%%%%%%%%%%%%%%%%%%%%%%%%%%%%%%%%%%%%%%%%%%%%%%%%
%%%%%%%%%%%%%%%%%%%%%%%%%%%%%%%%%%%%%%%%%%%%%%%%%%%%%%%
%%%%%%%%%%%%%%%%%%%%%%%%%%%%%%%%%%%%%%%%%%%%%%%%%%%%%%%
%%%%%%%%%%%%%%%%%%%%%%%%%%%%%%%%%%%%%%%%%%%%%%%%%%%%%%%
%%%%%%%%%%%%%%%%%%%%%%%%%%%%%%%%%%%%%%%%%%%%%%%%%%%%%%%
%%%%%%%%%%%%%%%%%%%%%%%%%%%%%%%%%%%%%%%%%%%%%%%%%%%%%%%
%%%%%%%%%%%%%%%%%%%%%%%%%%%%%%%%%%%%%%%%%%%%%%%%%%%%%%%
%%%%%%%%%%%%%%%%%%%%%%%%%%%%%%%%%%%%%%%%%%%%%%%%%%%%%%%
%%%%%%%%%%%%%%%%%%%%%%%%%%%%%%%%%%%%%%%%%%%%%%%%%%%%%%%
%%%%%%%%%%%%%%%%%%%%%%%%%%%%%%%%%%%%%%%%%%%%%%%%%%%%%%%
%%%%%%%%%%%%%%%%%%%%%%%%%%%%%%%%%%%%%%%%%%%%%%%%%%%%%%%
\end{document}